\documentstyle[12pt,a4wide]{article}
\newcommand{\news}{\setcounter{equation}{0}}

\def\eqn{\begin{equation}} 
\def\eeqn{\end{equation}}
\def\arr{\begin{array}} 
\def\earr{\end{array}}
\def\eqna{\begin{eqnarray}} 
\def\eeqna{\end{eqnarray}} 
\def\a{\alpha}
\def\b{\beta}

\def\w{\wedge} 
 
\def\O{\Omega} 
\def\e{\epsilon}
 
\def\m{\mu} 
\def\n{\nu} 
\def\la{\lambda} 

\def\t{\tau} 
 
\def\g{\gamma}

\begin{document}

\vspace*{-.6in}
\thispagestyle{empty}
\begin{flushright}
PUPT-1842\\
hep-th/9903128\\
\end{flushright}

\vspace{.3in}
{\Large 
\begin{center}
{\bf Intersecting D-branes and black holes \\ in type 0 string theory}
\end{center}}
\vspace{.3in}
\begin{center}
Miguel S. Costa\footnote{miguel@feynman.princeton.edu}\\
\vspace{.1in}
\emph{Joseph Henry Laboratories\\ Princeton University \\ Princeton, New Jersey 08544, USA}
\end{center}

\vspace{.5in}

\begin{abstract}
We study intersecting D-branes in type 0 string theories and show that the D$p_{\pm}$-brane
bound states obey similar intersecting rules as the D$p$-branes of the 
type II theories. The D$5_{\pm}$-D$1_{\pm}$ brane system is studied in detail. We show that 
the corresponding near-horizon geometry is the $AdS_3\times S^3\times T^4$ space and that 
there is no tachyon instability in this background. The Bekenstein-Hawking entropy is 
calculated. The worldvolume theory on the D$5_{\pm}$-D$1_{\pm}$ system is also studied. 
This theory contains both bosons and fermions and it is seen to arise as a projection of the
supersymmetric gauge theory related to the D5-D1 system of the type IIB theory. The 
Bekenstein-Hawking entropy formula is reproduced exactly using the dual conformal 
field theory.
\end{abstract}
\newpage

\section{Introduction}
\news
Recently there has been a renewed interest in the type 0 string theories 
\cite{DixonHarv,SeibWitt}. While in the type II theories the GSO projection acts
independently on the left- and right-moving sectors, in the type 0 theories this
projection ties both sectors together. More precisely, this so-called diagonal
GSO projection is just $( 1+(-1)^{F+\tilde{F}})/2$ in the ${\rm NS-NS}$ sectors
and $( 1\pm(-1)^{F+\tilde{F}})/2$ in the ${\rm R-R}$ sectors (the plus sign choice
gives the type 0B theory and the minus sign the type 0A theory). The resulting string 
theories have worldsheet supersymmetry but no space-time 
supersymmetry. The closed string spectrum is divided in the following sectors
(in the notation of \cite{Polc})
\eqn
\arr{ll}
{\rm 0A}: & ({\rm NS-,NS-}) \oplus ({\rm NS+,NS+}) 
\oplus ({\rm R+,R-}) \oplus ({\rm R-,R+})\ ,
\\
{\rm 0B}: & ({\rm NS-,NS-}) \oplus ({\rm NS+,NS+}) 
\oplus ({\rm R+,R+}) \oplus ({\rm R-,R-})\ .
\earr
\label{1.1}
\eeqn
Notice that these theories have no fermions in the bulk. The massless bosonic fields are 
similar to the type II theories with the ${\rm R-R}$ fields doubled. In addition, there is
a tachyon with $m^2=-2/\a'$ coming from the $({\rm NS-,NS-})$ sector.

The doubling of the ${\rm R-R}$ sector in the type 0 theories has some interesting 
consequences. For each $p$ (even or odd in the 0A or 0B theories, respectively) there are 
two $(p+1)$-form gauge potentials ${\cal A}_{p+1}$ and $\bar{\cal A}_{p+1}$ corresponding
to the ${\rm R-R}$ sectors in (\ref{1.1}). Thus, these theories will have two different types
of D$p$-branes and corresponding anti-branes \cite{KlebTsey}. It is convenient to redefine
the gauge fields and corresponding D-brane charges in the following way \cite{KlebTsey}
\eqn
\left( {\cal A}_{p+1}\right)_{\pm}=
\frac{1}{\sqrt{2}}\left( {\cal A}_{p+1}\pm \bar{\cal A}_{p+1}\right)\ ,
\ \ \ \ \ 
q_{\pm}=\frac{1}{2}\left( q\pm \bar{q}\right)\ .
\label{1.2}
\eeqn
In this notation we have D$p_{+}$-branes, D$p_{-}$-branes and the corresponding
anti-branes. A D$p_{+}$-brane carries the charges $q=1=\bar{q}$ and a 
D$p_{-}$-brane the charges $q=1=-\bar{q}$. In the type 0B theory and for $p=3$,
the self-dual field strength ${\cal F}_5=d{\cal A}_4$ and the anti-self-dual 
field strength $\bar{\cal F}_5=d\bar{\cal A}_4$ may be combined to form an unconstrained 
$5$-form field strength. Then the D$3_{+}$-brane will be electrically charged and the
D$3_{-}$-brane magnetically charged. 

Following a suggestion made by Polyakov \cite{Poly}, and the earlier work on the 
$AdS/CFT$ duality \cite{Mald,GKP,Witt}, Klebanov and Tseytlin \cite{KlebTsey} used
the type 0 string theory to study non-supersymmetric non-conformal gauge theories.
In particular, they considered the theory of $N$ coincident electrically charged 
D$3$-branes (D$3_{+}$-branes). The corresponding effective worldvolume theory 
is purely bosonic and tachyon free. More precisely, it is the four-dimensional 
$U(N)$ gauge theory with six adjoint scalars. It was then argued that the ${\rm R-R}$ flux 
in the type 0 dual background stabilizes the tachyon, as expected from the duality
equivalence. Since then a few papers reporting results on this model appeared
$[9-14]$ (see also \cite{FerrMart,AlvaGomez,AFS} for 
related work in non-critical strings). 

In a related development, Klebanov and Tseytlin \cite{KlebTsey2} also considered the 
theory of $N$ D$3_{+}$-branes coincident with $N$ D$3_{-}$-branes ($N$ coincident
D$3_{\pm}$-branes\footnote{A D$p_{\pm}$-brane is a bound state of a D$p_{+}$-brane and 
a D$p_{-}$-brane; a D$p_{\mp}$-brane is a bound state of a D$p_{+}$-brane and a anti 
D$p_{-}$-brane; similar definitions follow for the anti D$p_{\pm}$- and 
D$p_{\mp}$-branes.}). The novelty here is that the worldvolume 
theory for this bound state has fermions corresponding to open strings with
one end on a D$3_{+}$-brane and the other end on a D$3_{-}$-brane. This 
theory is the conformal $U(N)\times U(N)$ gauge theory  with six adjoint
scalars and eight bifundamental Weyl spinors (four in the $(N,\overline{N})$ plus 
four in the $(\overline{N},N)$). The dual type 0 background is the $AdS_5\times S^5$
space. Remarkably, the D$p_{\pm}$-brane bound states of the type 0 theories
share many properties of the BPS D$p$-branes of the type II theories. In 
particular, two of such parallel branes satisfy the no-force condition 
\cite{KlebTsey,KlebTsey2} (see also \cite{Zare,TseyZare} for details on the potentials
between different type 0 D$3$-branes).

The results described above lead us to consider the problem of intersecting D$p$-branes 
in the type 0 theories. We shall start in section two by studying
the forces between intersecting type 0 D-branes. We conclude that the 
D$p_{\pm}$-brane bound states may be intersected according to the same rules 
of the type II theories. In section three we consider the $D5_{\pm}$-$D1_{\pm}$
brane system and show that the corresponding near-horizon geometry is the 
$AdS_3\times S^3\times T^4$ space. In this background the tachyon field is
stabilized. We calculate the Bekenstein-Hawking entropy
in terms of the quantized D-brane charges. In section four we study the
worldvolume theory associated to the D$5_{\pm}$-D$1_{\pm}$ brane system. This theory
contains fermions corresponding to strings with one end on a D$p_{+}$-brane and
the other end on a D$p'_{-}$-brane.  Also, we show that this gauge theory is a 
$Z_2$ projection of the type II gauge theory for the 
D5-D1 brane system, i.e. the compactification to two dimensions of a $N=1$,
$D=6$ supersymmetric gauge theory \cite{Mald1}. By choosing an appropriate branch of our 
gauge theory, we define the dual conformal field theory and reproduce exactly the 
Bekenstein-Hawking entropy formula. We present our conclusions 
in section five.

\section{Intersecting D-branes in type 0 string theory}
\news

We shall start by reviewing the results presented in \cite{KlebTsey}
for the interacting energy between parallel D-branes. The type 0 D-branes were
earlier studied in \cite{BergGabe} by using the boundary state formalism (see also 
\cite{BianSagn,Sagn,Ange} for related work on the open string descendents of the type 0B 
theory). The cylinder amplitude between two likely charged parallel D$p$-branes (i.e. 
between two D$p_{+}$-branes or two D$p_{-}$-branes) is obtained by omitting the R sector
in the open string channel \cite{BergGabe,KlebTsey}. The result is
\eqn
A=V_{p+1}\int^{\infty}_{0}\frac{dt}{2t}\left( 8\pi^2\a't\right)^{-\frac{p+1}{2}}
e^{-\frac{tY^2}{2\pi\a'}}
\left(\left[\frac{f_3(q)}{f_1(q)}\right]^8-
\left[\frac{f_4(q)}{f_1(q)}\right]^8\right)\ ,
\label{2.1}
\eeqn
where $q=e^{-\pi t}$. The first term in the parenthesis comes from the NS spin structure
and the second term from the NS$(-1)^F$ spin structure. The open string tachyon is
projected out and the open string states are purely bosonic. In the closed string channel
the amplitude (\ref{2.1}) is seen to contain a term from the closed string tachyon, a
term from the attractive ${\rm NS-NS}$ massless exchange and a term from the repulsive 
${\rm R-R}$ massless exchange (which is twice the ${\rm NS-NS}$ attractive term because 
the ${\rm R-R}$ fields are doubled). 

To calculate the cylinder amplitude between a D$p_{+}$-brane and a D$p_{-}$-brane, the
correct projection in the open string channel is to retain the R sector only. The 
result is minus the amplitude in (\ref{2.1}). However, in this case the open strings are
associated with fermionic states. In the closed string channel we have a term from the
tachyon and an attractive term from the ${\rm NS-NS}$ massless exchange (the ${\rm R-R}$ 
contribution to the interacting potential cancels). 

A few interesting results follow from the previous analysis \cite{KlebTsey}. Firstly,
the worldvolume theory of a D$p_{\pm}$-brane bound state contains fermions arising from
strings ending on branes with different charges and bosons arising from strings ending
on likely charged branes. Secondly, the cylinder amplitude between two D$p_{\pm}$-branes
may be seen to vanish by adding the contributions from each constituent. Also, because
$q=0$ for the D$p_{\pm}$-brane, it decouples from the bulk tachyon, i.e. it is not
a source for such field (or $\bar{q}=0$ for the D$p_{\mp}$-brane).

Now we turn to the case of intersecting D-branes. We want to calculate the interacting
energy between a D$p$-brane and a D$p'$-brane in the type 0 theories. This problem is 
analogous to the type II case \cite{Lifs}. The open strings stretching between
these branes will have either DD, NN or ND($\equiv$ DN) boundary conditions. The only 
T-duality invariant quantity is the number of directions along which the open strings 
satisfy ND boundary conditions because the DD and NN boundary conditions are related by
T-duality. Taking into account that the ghost contribution will cancel two of the 
(DD or NN) coordinates we have $\sharp {\rm ND}=8-\sharp ({\rm NN+DD})$. If 
$\sharp {\rm ND}\ne 0$ there are fermionic zero modes in the NS$(-1)^F$ sector and
therefore this sector will not contribute (as well as the R$(-1)^F$ sector). 

To calculate the cylinder amplitude between $+$ (or $-$) charged D$p$- and D$p'$-branes 
we have to omit the R sector in the open string channel. The result is
\eqn 
A=V_{l+1}\int^{\infty}_{0}\frac{dt}{2t}\left( 8\pi^2\a't\right)^{-\frac{l+1}{2}}
e^{-\frac{tY^2}{2\pi\a'}}
\left[\frac{f_3(q)}{f_1(q)}\right]^{\sharp({\rm NN+DD})}
\left[\frac{f_2(q)}{f_4(q)}\right]^{\sharp {\rm ND}}\ ,
\label{2.2}
\eeqn
where the branes intersect over a $l$-brane. If $\sharp {\rm ND}=0$ (and therefore
$\sharp ({\rm NN+DD})=8$) we obtain the contribution from the NS spin structure
in the amplitude (\ref{2.1}) (to which the contribution from the NS$(-1)^F$ spin 
structure must be added). To factorize the cylinder in the closed string channel we need 
to consider the limit $t\rightarrow 0$. The result is
\eqn
\left[\frac{f_3(q)}{f_1(q)}\right]^{\sharp({\rm NN+DD})}
\left[\frac{f_2(q)}{f_4(q)}\right]^{\sharp {\rm ND}}
\sim
\frac{t^{\sharp ({\rm NN+DD})/2}}{2^{\sharp {\rm ND}/2}}
\left(e^{\pi/t}+(8-2\sharp {\rm ND})+{\cal O}(e^{-\pi/t})\right)\ .
\label{2.3}
\eeqn
The first term comes from the closed string tachyon and the second term from the 
${\rm NS-NS}$ massless exchange. Note that the absence of the NS$(-1)^F$ sector
reflects the fact that different D-branes do not interact through the ${\rm R-R}$
gauge fields \cite{Lifs}. Note also that the ${\rm NS-NS}$ massless exchange vanishes
for $\sharp {\rm ND}=4$ but the tachyonic term remains. 

The cylinder amplitude between a D$p_+$- and a D$p'_-$-brane is obtained by retaining
the open string R sector only. The result is
\eqn 
A=-V_{l+1}\int^{\infty}_{0}\frac{dt}{2t}\left( 8\pi^2\a't\right)^{-\frac{l+1}{2}}
e^{-\frac{tY^2}{2\pi\a'}}
\left[\frac{f_2(q)}{f_1(q)}\right]^{\sharp({\rm NN+DD})}
\left[\frac{f_3(q)}{f_4(q)}\right]^{\sharp {\rm ND}}\ .
\label{2.4}
\eeqn
Thus, the worldvolume theory of a D$p_+$- and a D$p'_-$-brane will contain fermions
on their intersection. For small $t$, we found
\eqn
-\left[\frac{f_2(q)}{f_1(q)}\right]^{\sharp({\rm NN+DD})}
\left[\frac{f_3(q)}{f_4(q)}\right]^{\sharp {\rm ND}}
\sim
\frac{t^{\sharp ({\rm NN+DD})/2}}{2^{\sharp {\rm ND}/2}}
\left(-e^{\pi/t}+(8-2\sharp {\rm ND})+{\cal O}(e^{-\pi/t})\right)\ .
\label{2.5}
\eeqn
Comparing with (\ref{2.3}) we see that the sign of the tachyonic term is reversed while
the term corresponding to the ${\rm NS-NS}$ massless exchange is the same. 

The potential between a D$p_{\pm}$- and a D$p'_{\pm}$-brane can be calculated by 
adding the potentials between the constituents. The result is twice the sum of
(\ref{2.2}) and (\ref{2.4}). In particular, for $\sharp {\rm ND}=4$ there is no 
force between these bound states. This is analogous to the type II case where for
$\sharp {\rm ND}=4$ we have a state preserving a fraction of the space-time
supersymmetry, i.e. a stable BPS state. In the type II case we are led to study the 
D$5$-D$1$ brane system that corresponds to a $5$-dimensional black hole in the 
dual gravitational description of the system \cite{StroVafa,CallMald}. Here we shall 
study the D$5_{\pm}$-D$1_{\pm}$ brane system of the type 0B theory. The worldvolume 
theory of such a bound state will contain fermions associated to open strings with
one end on a $+$ charged D-brane and the other end on a $-$ charged D-brane, and
bosons associated to strings ending on $+$ charged D-branes or on 
$-$ charged D-branes.

Finally, we consider the case $\sharp {\rm ND}=8$ which deserves a separate 
discussion\footnote{Recently, and while this work was in 
progress, a paper considering the intersection of a D$5_{+}$-brane and a D$5_{-}$-branes
over a string appeared \cite{BCR}. They used the existence of chiral fermions
on the intersection to derive the anomalous Wess-Zumino coupling \cite{GHM} for 
D$p_{+}$- and D$p_{-}$-branes.} \cite{Lifs}.
In this case there are no fermionic zero modes in the R$(-1)^F$ sector which has to be 
included. To calculate the amplitude for $+$ (or $-$) charged D$p$- and D$p'$-branes we 
retain the NS open string sector. The result is
\eqn 
A=V_{l+1}\int^{\infty}_{0}\frac{dt}{2t}\left( 8\pi^2\a't\right)^{-\frac{l+1}{2}}
e^{-\frac{tY^2}{2\pi\a'}}
\left[\frac{f_2(q)}{f_4(q)}\right]^8\ .
\label{2.6}
\eeqn
For a D$p_+$- and a D$p'_-$-brane we have the contributions from the R$(-1)^F$ and R
sectors with the result
\eqn 
A=V_{l+1}\int^{\infty}_{0}\frac{dt}{2t}\left( 8\pi^2\a't\right)^{-\frac{l+1}{2}}
e^{-\frac{tY^2}{2\pi\a'}}
\left(1-\left[\frac{f_3(q)}{f_4(q)}\right]^8\right)\ .
\label{2.7}
\eeqn
As in the type II theories, the cylinder amplitude between a D$p_{\pm}$- and a 
D$p'_{\pm}$-brane such that $\sharp {\rm ND}=8$ will vanish by the abstruse identity.

\section{The D$5_{\pm}$-D$1_{\pm}$ black hole solution}
\news

In the remainder of this paper we shall study the D$5_{\pm}$-D$1_{\pm}$ brane 
system. We start by considering the corresponding gravitational solution. The 
appropriate truncation of the type 0B effective action is \cite{KlebTsey}
\eqn
\arr{l}
S=\displaystyle{\frac{1}{2\kappa_{10}^2}\int d^{10}x\sqrt{-g}
\left[ e^{-2\phi}\left(R+4(\partial_a \phi)^{2}-\frac{1}{4}(\partial_a T)^{2}
-\frac{m^2}{4}T^2\right)\right.}
\\\\
\displaystyle{\left.\phantom{\left(\frac{m^2}{4}\right)}
\ \ \ \ \ \ \ \ \ \ \ \ \ \ \ \ \ \ \ \ \ 
-\frac{1}{2.3!}f_+(T)\left({\cal F}_{3+}\right)^{2}
-\frac{1}{2.3!}f_-(T)\left({\cal F}_{3-}\right)^{2}\right]}\ ,
\earr
\label{3.1}
\eeqn
where 
\eqn
f_{\pm}=1\pm T+{\cal O}(T^2)\ ,\ \ \ m^2=-2/\a'\ ,\ \ \ 
\kappa_{10}^2=(2\pi)^6\pi g^2\a'^4\ ,
\label{3.2}
\eeqn
and $a$ is a ten-dimensional index. The dilaton field $\phi$ has its zero 
mode $\phi_0$ subtracted.

We proceed by compactifying the above action to five dimensions with the 
ansatz \cite{CGKT}
\eqn
\arr{rcl}
ds^2&=&e^{\phi_6}\left[ e^{-\frac{2}{3}\la}ds_5^2+
e^{2\la}\left(dx_1+{\cal A}^{(K)}_{\m}dx^{\m}\right)^2\right]+e^{2\n}dx^Idx_I\ ,
\\
\phi&=&\phi_6+2\n\ ,
\earr
\label{3.3}
\eeqn
where $I,J,...=6,...,9$ and  $\m,\n,...$ are five-dimensional indices. The field $\phi_6$ 
is the dilaton field in the six-dimensional theory. We shall make the following ansatz for the
3-form field strengths ${\cal F}_{3\pm}$ and Ka\l u\.{z}a-Klein field ${\cal A}^{(K)}_{\m}$:
$\left({\cal F}_{3\pm}\right)_{\m\n 1}$ are electric corresponding to the D1-branes,
$\left({\cal F}_{3\pm}\right)_{\m\n\rho}$ are magnetic corresponding to the 
D5-branes and ${\cal A}^{(K)}_{\m}$ is electric corresponding to the 
Ka\l u\.{z}a-Klein modes along $x_1$. The action (\ref{3.1}) becomes
\eqn
\arr{l}
S=\displaystyle{\frac{1}{2\kappa_{5}^2}\int d^{5}x\sqrt{-g_5}
\left[ R-(\partial_{\m} \phi_6)^{2}-\frac{4}{3}(\partial_{\m} \la)^{2}
-4(\partial_{\m} \n)^{2}-\frac{1}{4}(\partial_{\m} T)^{2}
\right.}
\\\\
\ \ \ \ \ \ \ \ \ \ \ \ \ 
\displaystyle{-\frac{m^2}{4}e^{\phi_6-\frac{2}{3}\la}T^2
-\frac{1}{2.2!}e^{-\frac{4}{3}\la+4\n}
\left( f_+(T)\left({\cal F}_{+\m\n 1}\right)^{2}+
f_-(T)\left({\cal F}_{-\m\n 1}\right)^{2}\right)}
\\\\
\ \ \ \ \ \ \ \ \ \ \ \ \ 
\displaystyle{\left.
-\frac{1}{2.2!}e^{\frac{8}{3}\la}\left({\cal F}_{\m\n}^{(K)}\right)^2
-\frac{1}{2.3!}e^{\frac{4}{3}\la+4\n}
\left( f_+(T)\left({\cal F}_{+\m\n\rho}\right)^{2}+
f_-(T)\left({\cal F}_{-\m\n\rho}\right)^{2}\right)\right]}\ ,
\earr
\label{3.4}
\eeqn
where $\kappa_{5}^2=\kappa_{10}^2/V_5$ with $V_5$ the volume of the compact
five-dimensional space $T^5$. Comparing with the results in \cite{CGKT} the new
ingredients are the tachyon field and the doubling of the ${\rm R-R}$ fields.

Next we perform the following ansatz for the fields in the reduced action (\ref{3.4})
\eqn
\arr{l}
ds_5^2=-e^{2a(r)}dt^2+e^{2b(r)}dr^2+e^{2c(r)}d\O_3^2\ ,
\\\\
\phi_6=\phi_6(r)\ ,\ \ \ \la=\la(r)\ ,\ \ \ \n=\n(r)\ ,\ \ \ T=T(r)\ ,
\\\\
{\cal A}^{(K)}_t={\cal A}^{(K)}(r)\ ,\ \ \ 
{\cal A}_{\pm t1}={\cal A}_{\pm}(r)\ ,\ \ \ 
{\cal F}_{\pm \m\n\rho}=2Q_{5\pm}\e_{\m\n\rho}\ ,
\earr
\label{3.5}
\eeqn
where $\e$ is the unit 3-sphere volume form. It is easy to see that the magnetic ansatz
for ${\cal F}_{\pm \m\n\rho}$ automatically solves the corresponding equation of motion
and that the functions ${\cal A}^{(K)}(r)$ and ${\cal A}_{\pm}(r)$ satisfy
\eqn
\arr{rcl}
\displaystyle{\frac{d{\cal A}^{(K)}}{dr}}&=&2Q_Ke^{a+b-3c-\frac{8}{3}\la}\ ,
\\\\
\displaystyle{\frac{d{\cal A}_{\pm}}{dr}}&=&2Q_{1\pm}\left(f_{\pm}^{\phantom 2}(T)\right)^{-1}
e^{a+b-3c+\frac{4}{3}\la-4\n}\ .
\earr
\label{3.6}
\eeqn
The field equations for the remaining fields can be derived from an effective 
one-dimensional action $S=S[\phi_6,\la,\n,T,a,b,c]$. Performing the change of variable
$d\t=-2e^{-3c-a+b}dr$ and the field redefinitions \cite{CGKT}
\eqn
\rho=2c+a\ ,\ \ \ \a=a-\frac{4}{3}\la\ ,\ \ \ 
\b=a+\frac{2}{3}\la+2\n\ ,\ \ \ \g=a+\frac{2}{3}\la-2\n\ ,
\label{3.7}
\eeqn
this effective action takes the simple form
\eqn
S=\int d\t\left[ \frac{3}{2}\dot{\rho}^2-\frac{1}{2}\dot{\a}^2
-\frac{1}{2}\dot{\b}^2-\frac{1}{2}\dot{\g}^2-\dot{\phi_6}^2
-\frac{1}{4}\dot{T}^2-V(\rho,\a,\b,\g,\phi_6,T)\right]\ ,
\label{3.8}
\eeqn
where $\dot{}$ denotes derivative with respect to $\t$ and
\eqn
V=\frac{m^2}{16}T^2e^{3\rho+\phi_6-(\b+\g)/2}-\frac{3}{2}e^{2\rho}+
\frac{1}{2}Q_K^2e^{2\a}+h_1(T)e^{2\g}+h_5(T)e^{2\b}\ ,
\label{3.9}
\eeqn
with
\eqn
\arr{l}
h_1(T)=\displaystyle{
\frac{1}{2}\left(Q_{1+}^2\left(f_+^{\phantom 2}(T)\right)^{-1}+
Q_{1-}^2\left(f_-^{\phantom 2}(T)\right)^{-1}\right)}\ ,
\\\\
h_5(T)=\displaystyle{
\frac{1}{2}\left(Q_{5+}^2f_+(T)+Q_{5-}^2f_-(T)\right)}\ .
\earr
\label{3.9a}
\eeqn
The first term in the potential $V$ comes from the tachyon mass, the second
term from the curvature of $S^3$ and the remaining terms from the 
Ka\l u\.{z}a-Klein, D1-branes and D5-branes charges. The equations
of motion following from the action (\ref{3.8}) should be supplemented with the 
zero energy condition \cite{CGKT}
\eqn
\frac{3}{2}\dot{\rho}^2-\frac{1}{2}\dot{\a}^2
-\frac{1}{2}\dot{\b}^2-\frac{1}{2}\dot{\g}^2-\dot{\phi_6}^2
-\frac{1}{4}\dot{T}^2+V(\rho,\a,\b,\g,\phi_6,T)=0\ .
\eeqn

To find the solution to the above system first we consider the tachyon field.
For small $T$, we have $f_+(T)\sim 1/f_-(T)$ and therefore 
the function $h_1(T)$ has an extremum at $T=T_0$ such that $f_+(T_0)=Q_{1+}/Q_{1-}$ 
\cite{KlebTsey2}. Similarly, the function $h_5(T)$ has an extremum at $T=T_0$ such that 
$f_+(T_0)=Q_{5-}/Q_{5+}$. For $Q_{p+}=Q_{p-}\equiv Q_p$ ($p=1$ or $5$) we 
have $f_+(T_0)=1$ which is solved by $T_0=0$ and it is consistent with the assumption that
$T$ is small (more generally if $f_+=f_-^{-1}$ and $f_+(0)=1$ the 
argument will follow, e.g. for $f_{\pm}=e^{\pm T}$). Thus, the tachyon equation 
is solved by $T=0$. The solution for the dilaton field $\phi_6$ becomes
$\phi_6=0$ (recall that $\phi_6$ has its zero mode subtracted). Noting that
$h_p(0)=Q_p^2$, the solution is similar to the D5-D1 solution of the type IIB
theory obtained in \cite{CGKT} using the above method. In terms of the ten-dimensional
theory fields we have
\eqn
\arr{rcl}
ds_{10}^2&=&\displaystyle{
(H_1H_5)^{-\frac{1}{2}}\left(-dt^2+dx_1^2+\left(\frac{r_0}{r}\right)^2
\left(\cosh{(\b)}dt+\sinh{(\b)}dx_1^{\phantom 2}\right)^2\right)}
\\\\
&&\displaystyle{
+(H_1H_5)^{\frac{1}{2}}\left(dr^2+r^2d\O_3^2\right)+
H_1^{\frac{1}{2}}H_5^{-\frac{1}{2}}dx^Idx_I}\ ,
\\\\
e^{2\phi}&=&\displaystyle{\frac{H_1}{H_5}\ ,\ \ \ \ \ T\ =\ 0}\ ,
\\\\
{\cal F}_{3\pm}&=&\displaystyle{
\frac{2Q_1}{r^3}H_1^{-2}dt\w dx_1\w dr + 2Q_5\e}\ ,
\earr
\label{3.10}
\eeqn
where 
\eqn
\sqrt{2}Q_p=r_p^2\ ,\ \ \  Q_K=r_0^2\sinh{(2\b)}\equiv r_K^2\ ,\ \ \ 
H_p=1+\left(\frac{r_p}{r}\right)^2\ ,\ \ \ h=1-\left(\frac{r_0}{r}\right)^2\ . 
\eeqn
Notice that we are considering the dilute gas regime $r_p\gg r_0,r_K$.
The fields $(\a,\b,\g,\rho)$ corresponding to the solution (\ref{3.10}) are 
\eqn
e^{\a}=h^{\frac{1}{2}}H_K^{-1}\ ,\ \ \ 
e^{\b}=h^{\frac{1}{2}}H_5^{-1}\ ,\ \ \ 
e^{\g}=h^{\frac{1}{2}}H_1^{-1}\ ,\ \ \ 
e^{\rho}=r^2h^{\frac{1}{2}}\ . 
\label{3.11}
\eeqn

Next we discuss the issue of stability of the background (\ref{3.10}). We consider small
perturbations of the tachyon field $T$ around $T_0=0$. The leading quadratic terms in the 
potential $V$ are seen to be
\eqn
V_2=h\left[\frac{m^2}{8}r^6(H_1H_5)^{\frac{1}{2}}+
Q_1^2\left(f'_+(0)\right)^2H_1^{-2}+
Q_5^2\left(f'_+(0)\right)^2H_5^{-2}\right]\frac{T^2}{2}+{\cal O}(T^3)\ .
\label{3.12}
\eeqn
Asymptotically the background (\ref{3.10}) is just Minkowsky space which is not stable
because of the tachyon field. However, we are interested in the near-horizon 
geometry where we expect the tachyonic instability to be cured. The near-horizon region
corresponds to the limit $r\rightarrow r_0$ such that the dilute gas condition 
$r_p\gg r_0,r_K,r$ holds. In this limit the metric in (\ref{3.10}) becomes 
\eqn
\arr{rcl}
ds_{10}^2&=&\displaystyle{\frac{r^2}{r_1r_5}\left(-dt^2+dx_1^2\right)
+\frac{r_0^2}{r_1r_5}\left(\cosh{(\b)}dt+\sinh{(\b)}dx_1^{\phantom 2}\right)^2}
\\\\
&&\displaystyle{+\frac{r_1r_5}{r^2-r_0^2}dr^2+r_1r_5d\O_3^2+\frac{r_1}{r_5}dx^Idx_I}\ .
\earr
\label{3.13}
\eeqn
This geometry is the product of the BTZ black hole which is locally $AdS_3$ and
$S^3\times T^4$. In this background the potential $V_2$ becomes
\eqn
V_2=\left( r^2-r_0^2\right)r_0^2\left[\frac{m^2}{8}r_1r_5+
\left(f_+'(0)\right)^2\right]\frac{T^2}{2}+{\cal O}(T^3)\ .
\label{3.14}
\eeqn
Thus, the tachyon is stabilized for $r_1r_5\le 4\a'$.\footnote{Note that in a $AdS$
background tachyons with sufficiently small negative mass square are allowed
\cite{BreiFree}. However, this fact does not change the above condition 
significantly \cite{KlebTsey2}.} But this is precisely the region where
the $\a'$ corrections to the gravity approximation are expected to be important. However,
because the $AdS_3$ and $S^3$ radii are equal the Weyl tensor for this background cancels. 
Assuming that the $\a'$ corrections can be written in terms of the Weyl tensor we expect 
this geometry to be an exact solution of the theory \cite{GKT}. Also, the stability 
condition for the tachyon field is expected to hold \cite{KlebTsey2}.

Next we quantize the parameters in the gravitational solution. This can be done by
relating the quanta of D$5_{\pm}$-brane, D$1_{\pm}$-brane and Ka\l u\.{z}a-Klein charge
with the parameters $r_5$, $r_1$, $r_0$ and $\b$. The result is
\eqn
\arr{l}
\displaystyle{N_5=N_{5\pm}=\frac{1}{4\pi^2g\a'}\int_{S^3}{\cal F}_{3\pm}=
\frac{r_5^2}{\sqrt{2}g\a'}}\ ,
\\\\
\displaystyle{N_1=N_{1\pm}=\frac{v}{4\pi^2g\a'}\int_{S^3\times T^4}\star{\cal F}_{3\pm}=
\frac{vr_1^2}{\sqrt{2}g\a'}}\ ,
\\\\
\displaystyle{N_{L,R}=\frac{R_1^2v}{4g^2\a'^2}r_0^2e^{\pm2\b}}\ ,
\earr
\label{3.15}
\eeqn
where $v=(R_6...R_9)/\a'^2$. Comparing with the type II case we have an extra factor of
$1/\sqrt{2}$ in the expressions for $N_5$ and $N_1$. This factor arises because the type 0
D$p$-brane tension $T_p$ is related to the type II D$p$-brane tension $T_p^{II}$ by 
$T_p=T_p^{II}/\sqrt{2}$ and because the parameter $r_p$ is related to the mass of $N_p$
D$p_+$-branes and $N_p$ D$p_-$-branes (more precisely we have
$\frac{2\pi^2}{\kappa_5^2}\frac{r_p^2}{2}=N_p\frac{T_p^{II}}{\sqrt{2}}V_p$).
We remark that the six-dimensional string
coupling is $g_6=g/\sqrt{v}$ and that the condition to stabilize the tachyon derived above
can be written as $g_6\sqrt{N_1N_5}\le 2\sqrt{2}$.

We are now in position to calculate the Bekenstein-Hawking entropy in terms of the 
quantized charges. The result is
\eqn
S_{BH}=\frac{4\pi^3}{\kappa_5^2}r_1r_5\cosh{(\b)}=2\pi\sqrt{2N_1N_5}
\left(\sqrt{N_L}+\sqrt{N_R}\right)\ .
\label{3.16}
\eeqn
Notice that there is an extra factor of $\sqrt{2}$ in comparison with the type
IIB case.

\section{The D$5_{\pm}$-D$1_{\pm}$ worldvolume theory}
\news

In this section we study the worldvolume theory of the D$5_{\pm}$-D$1_{\pm}$ brane
system. Our main goal is to derive the entropy formula (\ref{3.16}) by considering a
specific branch of this theory. We shall see that the field theory for the 
D$5_{\pm}$-D$1_{\pm}$ system is a projection of the type IIB field theory for the D5-D1 
brane system. 

We start by analyzing the worldvolume field content on the D$5_{\pm}$-D$1_{\pm}$ bound
state. However, it is convenient to study instead the field theory on the T-dual 
D$9_{\pm}$-D$5_{\pm}$ bound state. As explained in section two, strings with both ends on 
$+$ charged or $-$ charged D-branes give worldvolume bosons while strings with one end on 
a $+$ charged D-brane and the other end on a $-$ charged D-brane give worldvolume fermions.
Schematically we may write
\eqn
\arr{rcl}
{\rm Bosons}&:&
\left(5_{\pm},5_{\pm}\right);\ \left(9_{\pm},9_{\pm}\right);\ 
\left(5_{\pm},9_{\pm}\right);\ \left(9_{\pm},5_{\pm}\right)\ ,
\\
{\rm Fermions}&:&
\left(5_{\pm},5_{\mp}\right);\ \left(9_{\pm},9_{\mp}\right);\ 
\left(5_{\pm},9_{\mp}\right);\ \left(9_{\pm},5_{\mp}\right)\ ,
\earr
\label{4.1}
\eeqn
where a $(5_+,5_+)$ string represents a string with both ends on a D$5_+$-brane and so on.
The corresponding massless modes that determine the field content of this gauge theory 
are\footnote{The subscripts $+$ and $-$ in the gauge groups allow us to distinguish between the 
$U(N)$ factors corresponding to $+$ or $-$ charged branes.}:
\begin{description}
\item[$\bullet$] $(5_{\pm},5_{\pm})$ strings: One gauge vector $A_{\a}^{(5,5)}$ 
$(\a=0,1,...,5)$ and four scalar fields $\phi^I_{(5,5)}$ $(I=6,...,9)$ in the
adjoint representation of $U_+(N_1)\times U_-(N_1)$.
\item[$\bullet$] $(9_{\pm},9_{\pm})$ strings: One gauge vector $A_{\a}^{(9,9)}$ 
and four scalar fields $\phi^I_{(9,9)}$ in the
adjoint representation of $U_+(N_5)\times U_-(N_5)$. 
\item[$\bullet$] $(5_+,9_+)$ and $(9_+,5_+)$ strings: One $SU(2)$ doublet $\chi$ describing
two complex scalars in the $(N_1,\overline{N}_5)$ of $U_+(N_1)\times U_+(N_5)$.
\item[$\bullet$] $(5_-,9_-)$ and $(9_-,5_-)$ strings: One $SU(2)$ doublet $\chi'$ describing
two complex scalars in the $(N_1,\overline{N}_5)$ of $U_-(N_1)\times U_-(N_5)$.
\item[$\bullet$] $(5_+,5_-)$ strings: Two Weyl spinors $\Psi_i^{(5,5)}$ $(i=1,2)$ in
the $(N_1,\overline{N}_1)$ of $U_+(N_1)\times U_-(N_1)$.
\item[$\bullet$] $(5_-,5_+)$ strings: Two Weyl spinors $\Psi'^{(5,5)}_i$ in
the $(\overline{N}_1,N_1)$ of $U_+(N_1)\times U_-(N_1)$.
\item[$\bullet$] $(9_+,9_-)$ strings: Two Weyl spinors $\Psi_i^{(9,9)}$ in
the $(N_5,\overline{N}_5)$ of $U_+(N_5)\times U_-(N_5)$.
\item[$\bullet$] $(9_-,9_+)$ strings: Two Weyl spinors $\Psi'^{(9,9)}_i$ in
the $(\overline{N}_5,N_5)$ of $U_+(N_5)\times U_-(N_5)$.
\item[$\bullet$] $(5_+,9_-)$ and $(9_-,5_+)$ strings: One Weyl spinor $\la$ in
the $(N_1,\overline{N}_5)$ of $U_+(N_1)\times U_-(N_5)$.
\item[$\bullet$] $(5_-,9_+)$ and $(9_+,5_-)$ strings: One Weyl spinor $\la'$ in
the $(N_1,\overline{N}_5)$ of $U_-(N_1)\times U_+(N_5)$.
\end{description}

Now we show that the gauge theory for the D$9_{\pm}$-D$5_{\pm}$ system can be obtained
as a projection of the analogous supersymmetric gauge theory of the type IIB theory, i.e. 
a $N=1$, $D=6$ supersymmetric gauge theory. We start with two vectormultiplets in the
adjoint of $U(2N_1)$ and $U(2N_5)$ together with two hypermultiplets in the same 
representation of the gauge groups. These fields can be written as a vector gauge field,
two Weyl spinors and four scalars, i.e. $(A_{\a}^{(5,5)},\ \Psi_i^{(5,5)},\ \phi^I_{(5,5)})$
and $(A_{\a}^{(9,9)},\ \Psi_i^{(9,9)},\ \phi^I_{(9,9)})$ in the adjoint of $U(2N_1)$ and
$U(2N_5)$, respectively. This is the field content of the $N=2$, $D=6$ supersymmetric
gauge theory that arises from the reduction on $T^4$ of the $N=1$, $D=10$ super 
Yang-Mills theory. The theory has another hypermultiplet $(\chi,\la)$ where 
$\chi$ is a $SU(2)$ doublet describing two complex scalars and $\la$ a Weyl
spinor both transforming in the $(2N_1,\overline{2N}_5)$ of $U(2N_1)\times U(2N_5)$.

The D$9_{\pm}$-D$5_{\pm}$ field theory is obtained by a $Z_2$ projection of the above 
theory. We take the $Z_2$ action to be $(-1)^F{\cal I}$ where $F$ is the fermionic 
number and ${\cal I}$ is the conjugation by $Diag(I,-I)$ where $I$ is
the $N_1\times N_1$ or the $N_5\times N_5$ identity matrix \cite{KlebTsey2}. This
projection breaks the gauge symmetry $U(2N_1)$ to $U_+(N_1)\times U_-(N_1)$ and
$U(2N_5)$ to $U_+(N_5)\times U_-(N_5)$. The action of this projection on the fields
$(A_{\a}^{(n,n)},\ \Psi_i^{(n,n)},\ \phi^I_{(n,n)})$ $(n=5,9)$ gives the fields
$(A_{\a}^{(n,n)},\ \Psi_i^{(n,n)},\ \Psi_i'^{(n,n)},\ \phi^I_{(n,n)})$ of our
theory and it is similar to the D$3_{\pm}$-brane case discussed in \cite{KlebTsey2}.
To determine the action on the hypermultiplet $(\chi,\la)$ we decompose these
fields in $N_1\times N_5$ blocks. The conjugation by ${\cal I}$
\eqn
\left(
\arr{cc}
A&B\\
C&D
\earr
\right)\rightarrow
\left(
\arr{cc}
I_{N_1}&0\\
0&-I_{N_1}
\earr
\right)
\left(
\arr{cc}
A&B\\
C&D
\earr
\right)
\left(
\arr{cc}
I_{N_5}&0\\
0&-I_{N_5}
\earr
\right)=
\left(
\arr{cc}
A&-B\\
-C&D
\earr
\right)\ ,
\label{4.3}
\eeqn
together with the $(-1)^F$ action leaves us with the fields $\chi,\chi'$ and $\la,\la'$
previously discussed.

To calculate the entropy associated with the gravitational background (\ref{3.13})
we assume that the dynamics of this black hole is determined by the strings 
stretching between D$1_{\pm}$- and D$5_{\pm}$-branes.
This is analogous to what happens in the type IIB case. Thus, the system is described
by $2\times(4N_1N_5)$ massless bosons arising from the fields $\chi$ and $\chi'$ and
by $2\times(4N_1N_5)$ massless fermions arising from the fields $\la$ and $\la'$, where
we are considering the reduction of our gauge theory to two dimensions. Hence, we have 
a conformal field theory with central charge 
$c=2\times(4+2)N_1N_5$. This effective theory has to be conformal because of the $AdS_3$
character of the dual type 0B background. If we excite this system with $N_{L,R}$ units
of left- and right-moving momentum, we have the usual asymptotic formula for the 
entropy
\eqn
S=\pi\sqrt{c\frac{2}{3}}\left(\sqrt{N_L}+\sqrt{N_R}\right)=
2\pi\sqrt{2N_1N_5}\left(\sqrt{N_L}+\sqrt{N_R}\right)\ ,
\label{4.2}
\eeqn
which is valid for $N_{L,R}\gg N_1N_5$ and exactly reproduces the result (\ref{3.16}).
The previous condition may be relaxed by considering winding D-branes \cite{MaldSuss}.

Finally, we justify the assumption that the black hole dynamics is determined by open 
strings stretching between D$1_{\pm}$- and D$5_{\pm}$-branes. We assume that the dual 
field theory is the projection $(-1)^F{\cal I}$ of the dual field theory in the type II
case, i.e. the Higgs branch of the $N=1$, $D=6$ supersymmetric gauge theory (compactified
on $T^4$) discussed above. This branch corresponds to exciting the hypermultiplets while the 
vectormultiplets are in their vacuum state. To analyze the projected theory we consider 
for simplicity only the bosonic degrees of freedom. The D-terms of the original 
supersymmetric theory become in the projected theory
\eqn
\arr{l}
D^a_{IJ}=f^a_{bc}\left( \phi^b_I\phi^c_J+\frac{1}{2}\e_{IJKL}\phi^b_K\phi^c_L\right)
+\chi^{\dagger}T^a\Gamma_{IJ}\chi\ ,
\\\\
D^{a'}_{IJ}=f^{a'}_{b'c'}\left( \phi^{b'}_I\phi^{c'}_J+
\frac{1}{2}\e_{IJKL}\phi^{b'}_K\phi^{c'}_L\right)
+\chi'^{\dagger}T^{a'}\Gamma_{IJ}\chi'\ ,
\earr
\label{4.4}
\eeqn
where we are using the notation in \cite{Mald1}. The indices $a,b,c$ run over the gauge 
group generators of $U_+(N_1)$ and $U_+(N_5)$ and the indices $a',b',c'$ of 
$U_-(N_1)$ and $U_-(N_5)$. To define a vacuum of this theory we must set the potential 
\eqn
V=\sum_{aIJ}\left( D^a_{IJ}\right)^2+\sum_{a'IJ}\left( D^{a'}_{IJ}\right)^2\ ,
\label{4.5}
\eeqn
to zero. This gives $2\times(3N_1^2)+2\times(3N_5^2)$ conditions. The number of
independent gauge transformations is $2N_1^2+2N_5^2$. Since the number of massless 
bosons in the projected Higgs branch is $2\times(4N_1N_5+4N_1^2+4N_5^2)$ we are
left with $2\times(4N_1N_5)$ bosons parameterizing the moduli space of vacua. But
these are precisely the number of bosons associated with the open strings stretching
between D$1_{\pm}$- and D$5_{\pm}$-branes.

It is interesting to relate the above description to the instanton moduli
space approximation of \cite{StroVafa}. In the latter description we view the 
D$1_{\pm}$-branes as instantons in the D$5_{\pm}$-branes gauge theory. This 
theory is the $U_+(N_5)\times U_-(N_5)$ six-dimensional gauge theory with four
adjoint scalars and four bifundamental Weyl spinors (two in the 
$(N_5,\overline{N}_5)$ and two in the $(\overline{N}_5,N_5)$). The $N_1$
D$1_{\pm}$-branes are seen as instantons in this gauge theory with instanton 
number $\n=N_{1+}+N_{1-}=2N_1$. Our results predict that the moduli space of these
instantons is given by a conformal field theory with $2\times(4N_1N_5)$ species
of bosons and $2\times(4N_1N_5)$ species of fermions, i.e. with central charge 
$c=12N_1N_5$.

\section{Conclusion}

Let us start by summarizing the results presented in this paper. We started by 
considering intersecting D-branes in type 0 string theory. We calculated the
cylinder amplitudes for intersecting D$p$- and D$p'$-branes with the result 
that open strings terminating on $+$ (or $-$) charged D-branes give worldvolume 
bosons while open strings terminating on a $+$ charged D-brane and on a $-$ 
charged D-brane give worldvolume fermions.
We concluded that the intersecting rules for the D$p_{\pm}$-brane bound states are identical 
to the intersecting rules for the D-branes of the type II theories. In the remainder
of this paper we analyzed in detail the D$5_{\pm}$-D$1_{\pm}$ brane system of the type
0B theory. We showed that the near-horizon geometry for this D-brane bound state is the 
$AdS_3\times S^3\times T^4$ space (the $AdS_3$ factor may be replaced by the BTZ black 
hole) which was argued to be a stable and exact solution of the type 0B gravitational 
equations. We proceeded by describing the worldvolume field theory on this 
D-brane system. It turned out that this gauge theory is a $Z_2$ projection of the
analogous theory for the D5-D1 brane system of the type IIB theory. By projecting
the Higgs branch of the latter theory we correctly reproduced the Bekenstein-Hawking
entropy formula calculated in the gravitational analysis. Note that this agreement 
between the gravitational and microscopic entropy calculations may be seen to arise
because the type 0B theory is a orbifold of the type IIB theory. As a consequence,
the gravitational solution is similar to the type IIB case but with the doubling
of the ${\rm R-R}$ fields, while the gauge theory is obtained by a $Z_2$ projection
of the type IIB D5-D1 gauge theory.

Finally, we note that because the rules for intersecting the D$p_{\pm}$-brane
bound states are the same as for the type II theories, there are many results that 
are expected to hold in the type 0 theories. In particular, we expect to have $D=4$
black holes corresponding to other intersecting D$p_{\pm}$-branes, e.g. four 
intersecting D$3_{\pm}$-branes \cite{KlebTsey3}. 
Also, we expect to have rotating black holes \cite{CvetLars}. 
Furthermore, we expect that more generally the 
D$p_{\pm}$-branes can intersect at $SU(N)$ angles \cite{BDL}. The corresponding
gravitational background can be derived by using the results in refs. \cite{BMM,
BLL,CostaCvet}. Also, there should be bound states of branes with constant worldvolume
magnetic fields and corresponding black hole solutions \cite{CostaPerry}.

\section*{Acknowledgments}

I would like to thank Michael Gutperle and Lori Paniak for discussions
and specially Igor Klebanov for very useful discussions and for reading a draft of the
paper. This work was supported by FCT (Portugal) under programme PRAXIS XXI and
by the NSF grant PHY-9802484.


\newpage

\begin{thebibliography}{40}
\bibitem{DixonHarv}L. Dixon and J. Harvey, Nucl. Phys. {\bf B274} (1986) 93.
\bibitem{SeibWitt}N. Seiberg and E. Witten, Nucl. Phys. {\bf B276} (1986) 272.
\bibitem{Polc}J. Polchinski, {\em String Theory}, Vol. 2, 
Cambridge University Press, 1998.
\bibitem{KlebTsey}I.R. Klebanov and A.A. Tseytlin, {\em D-Branes and Dual Gauge Theories 
in Type 0 Strings}, hep-th/9811035.
\bibitem{Poly}A.M. Polyakov, {\em The Wall of the Cave}, hep-th/9809057.
\bibitem{Mald}J.M. Maldacena, Adv. Theor. Math. Phys. {\bf 2} (1998) 231.
\bibitem{GKP}S.S. Gubser, I.R. Klebanov and A.M. Polyakov, Phys. Lett. 
{\bf B428} (1998) 105. 
\bibitem{Witt}E. Witten, Adv. Theor. Math. Phys. {\bf 2} (1998) 253.
\bibitem{Mina}J. Minahan, {\em Glueball Mass Spectra and Other Issues for 
Supergravity Duals of QCD Models}, hep-th/9811156.
\bibitem{KlebTsey1}I.R. Klebanov and A.A. Tseytlin, {\em Asymptotic Freedom and Infrared
Behavior in the Type 0 String Approach to Gauge Theory}, hep-th/9812089.
\bibitem{Garo}M.R. Garousi, {\em String Scattering from D-branes in Type 0 Theories},
hep-th/9901085.
\bibitem{Zare}K. Zarembo, {\em Coleman-Weinberg Mechanism and Interaction of
D3-Branes in Type 0 String Theory}, hep-th/9901106.
\bibitem{Mina1}J. Minahan, {\em Asymptotic Freedom and Confinement from Type 0 
String Theory}, hep-th/9902074. 
\bibitem{KoganLuzon}I.I. Kogan and G. Luz\'on, {\em Scale Invariance of Dirac 
Condition $g_e g_m = 1$ in Type 0 String Approach to Gauge Theory}, hep-th/9902086.
\bibitem{FerrMart}G. Ferretti and D. Martelli, {\em On the construction of gauge 
theories from non critical type 0 strings}, hep-th/9811208.
\bibitem{AlvaGomez}E. Alvarez and C. Gomez, {\em Non-Critical Confining Strings 
and the Renormalization Group}, hep-th/9902012.
\bibitem{AFS}A. Armoni, E. Fuchs and J. Sonnenschein, {\em Confinement in 4D Yang-Mills 
Theories from Non-Critical Type 0 String Theory}, hep-th/9903090.
\bibitem{KlebTsey2}I.R. Klebanov and A.A. Tseytlin, {\em A Non-supersymmetric Large
N CFT from Type 0 String Theory}, hep-th/9901101.
\bibitem{TseyZare}A.A. Tseytlin and K. Zarembo, {\em  Effective potential in 
non-supersymmetric $SU(N)\times SU(N)$ gauge theory and interactions of type 
0 D3-branes}, hep-th/9902095.
\bibitem{Mald1}J.M. Maldacena, {\em Black Holes in String Theory},
Ph.D. Thesis, hep-th/9607235.
\bibitem{BergGabe}O. Bergman and M.R. Gaberdiel, Nucl. Phys. {\bf B499} (1997) 183.
\bibitem{BianSagn}M. Bianchi and A. Sagnotti, Phys. Lett. {\bf B247} (1990) 517.
\bibitem{Sagn}A. Sagnotti, {\em Some Properties of Open - String Theories}, hep-th/9509080;
Nucl. Phys. Proc. Suppl. {\bf B56} (1997) 332. 
\bibitem{Ange}C. Angelantonj, Phys. Lett. {\bf B444} (1998) 309.
\bibitem{Lifs}G. Lifschytz, Phys. Lett. {\bf B388} (1996) 720. 
\bibitem{StroVafa}A. Strominger and C. Vafa, Phys. Lett. {\bf B379} (1996) 99.
\bibitem{CallMald}C. Callan and J.M. Maldacena, Nucl. Phys. {\bf B472} (1996) 591.
\bibitem{BCR}M. Bill\'o, B. Craps and F. Roose, {\em On D-branes in Type 0 String Theory},
hep-th/9902196.
\bibitem{GHM}M. Green, J. Harvey and G. Moore, Class. Quant. Grav. {\bf 14} (1997) 47.
\bibitem{CGKT}C.G. Callan, S.S. Gubser, I.R. Klebanov and A.A. Tseytlin,
Nucl. Phys. {\bf B489} (1997) 65.
\bibitem{BreiFree}P. Breitenlohner and D.Z. Freedman, Ann. Phys. {\bf 144} (1982) 249.
\bibitem{GKT}S.S. Gubser, I.R. Klebanov and A.A. Tseytlin, Nucl.Phys. {\bf B534} (1998) 202.  
\bibitem{MaldSuss}J.M. Maldacena and L. Susskind, Nucl. Phys. {\bf B475}
(1996) 679.
\bibitem{KlebTsey3}I.R. Klebanov and A.A. Tseytlin, Nucl.Phys. {\bf B475} (1996) 179.
\bibitem{CvetLars}M. Cveti\v{c} and F. Larsen, Nucl. Phys. {\bf B531} (1998) 239; 
{\em Microstates of Four-Dimensional Rotating Black Holes from 
Near-Horizon Geometry}, hep-th/9805146.
\bibitem{BDL}M. Berkooz, M.R. Douglas and R.G. Leigh, Nucl. Phys. {\bf B480} (1996) 265.
\bibitem{BMM}J.C. Breckenridge, G. Michaud and R.C. Myers,
Phys. Rev. {\bf D56} (1997) 5172.
\bibitem{BLL}V. Balasubramanian, F. Larsen, R.G. Leigh,
Phys. Rev. {\bf D57} (1998) 3509.
\bibitem{CostaCvet}M.S. Costa and M. Cveti\v{c}, Phys. Rev. {\bf D56} (1997) 4834-4843.
\bibitem{CostaPerry}M.S. Costa and M.J. Perry, Nucl. Phys. {\bf B520} (1998) 205;
Nucl. Phys. {\bf B524} (1998) 333; M.S. Costa, J. High Energy Phys. {\bf 11} (1998) 007.
\end{thebibliography}
\end{document}